\documentclass[12pt,a4paper,dvips,twoside]{article}
\usepackage{color,graphicx,psfig}
\usepackage{cite}

\definecolor{dark-blue}{rgb}{0,0,0.5}

\begin{document}

{\bf \Large \sf Quantum Phase Transitions in Models of Magnetic Impurities}

R. Bulla, M. Vojta\\
Theoretische Physik III, Elektronische Korrelationen und
Magnetismus,\\ Universit\"at Augsburg, 86135 Augsburg, Germany

\vspace*{0.1cm}

\begin{quote}
Zero temperature phase transitions not only occur in the
bulk of quantum systems, but also at boundaries or impurities.
We review recent work on quantum phase transitions in impurity
models that are generalizations of the standard Kondo model
describing the interaction of a localized magnetic
moment with a metallic fermionic host.
Whereas in the standard case the moment is screened for
any antiferromagnetic Kondo coupling as $T\to 0$,
the common feature of all systems considered here is that
Kondo screening is suppressed due to the competition with
other processes.
This competition
can generate unstable fixed points associated with phase transitions,
where the impurity properties undergo qualitative changes.
In particular, we discuss the coupling to both non-trivial
fermionic and bosonic baths as well as two-impurity models,
and make connections to recent experiments.
\end{quote}
\section{Introduction}

Quantum mechanical systems can undergo zero-temperature phase
transitions upon variation of a non-thermal control parameter \cite{Sachdev},
where order is destroyed solely by quantum fluctuations.
Quantum phase transitions occur as a result of competing ground
state phases, and can be classified into first-order and continuous
transitions.
The transition point of a continuous quantum phase transition,
the so-called quantum-critical point, is typically characterized
by a critical continuum of excitations, and can lead to
unconventional behavior -- such as non-trivial power laws
or non-Fermi liquid physics - over a wide range of the phase
diagram (see Fig. 1a).

An interesting class of quantum phase transitions are so-called
{\em boundary} transitions where only the degrees of freedom of a subsystem
become critical.
In this paper we consider impurity transitions --
the impurity can be understood as a zero-dimensional boundary --
where the {\em impurity} contribution to the free energy
becomes singular at the quantum critical point.
Such impurity quantum phase transitions require the
thermodynamic limit in the bath system, but are completely
independent of possible {\em bulk} phase transitions in the
bath.

Our model systems are built from magnetic moments which
 show the Kondo effect \cite{Hewson}.
Originally, this effect describes the behavior of localized
magnetic impurities in metals. The relevant microscopic models
are the Kondo model and the single-impurity Anderson
model.
In the standard case (i.e. a single magnetic
impurity with spin $\frac{1}{2}$ coupling to a single conduction
band with a finite density of states (DOS) near the
Fermi level) screening of the magnetic
moment occurs below a temperature scale $T_{\rm K}$.
The screening is associated with the flow to strong coupling
of the effective interaction between impurity and host fermions.
The Kondo temperature $T_{\rm K}$
depends exponentially on the system parameters.


In this paper we want to give a brief summary of
impurity models
where the flow to strong coupling is prevented by the competition
with other processes, and, depending on the system
parameters, weak and/or intermediate-coupling
fixed points can be reached.
We are particularly interested in intermediate-coupling
fixed points, which typically show a finite ground-state entropy
and are unstable, which means that to reach them
requires either fine tuning of the couplings (typically associated
with an impurity phase transition) or the presence of special
symmetries (e.g., channel symmetry in multi-channel models).
In the following sections we discuss single-impurity models
with a coupling to a non-trivial fermionic bath (Sec. 2),
to multiple fermionic baths (Sec. 3), to both fermionic and bosonic baths
(Sec. 4), and finally systems of coupled magnetic impurities
(Sec. 5).
In quoting results, we refer to perturbative renormalization group (RG)
treatments, to numerical RG (NRG) calculations, and to exact results if
available.
We restrict ourselves to models with spin-$\frac{1}{2}$ impurities.


\section{Pseudogap Kondo and Anderson models}

A straightforward possibility to suppress Kondo screening is
to reduce the electron bath DOS at the Fermi level to zero --
 in a superconductor this can also be interpreted as
competition between Kondo effect and superconducting pairing.
Two cases have to be distinguished: so-called hard-gap and
pseudogap (soft-gap) systems.

In the hard-gap case, the DOS $\rho(\varepsilon)$ is zero in a
finite energy interval around the Fermi level.
Then, the absence of low-energy states prevents
screening for gap values exceeding the energy gain
due to Kondo screening.
The resulting transition between a local-moment (LM) phase without Kondo
screening, realized at small Kondo coupling $J$, and a screened
strong-coupling (SC) phase, reached for large $J$, is of first order,
and it occurs only in the presence of particle-hole (p-h) asymmetry.
In the p-h symmetric case, the local-moment state
persists for arbitrary values of the coupling~\cite{hardgap}.

\begin{figure}
\label{one}
\centerline{
\includegraphics[width=6.1cm]{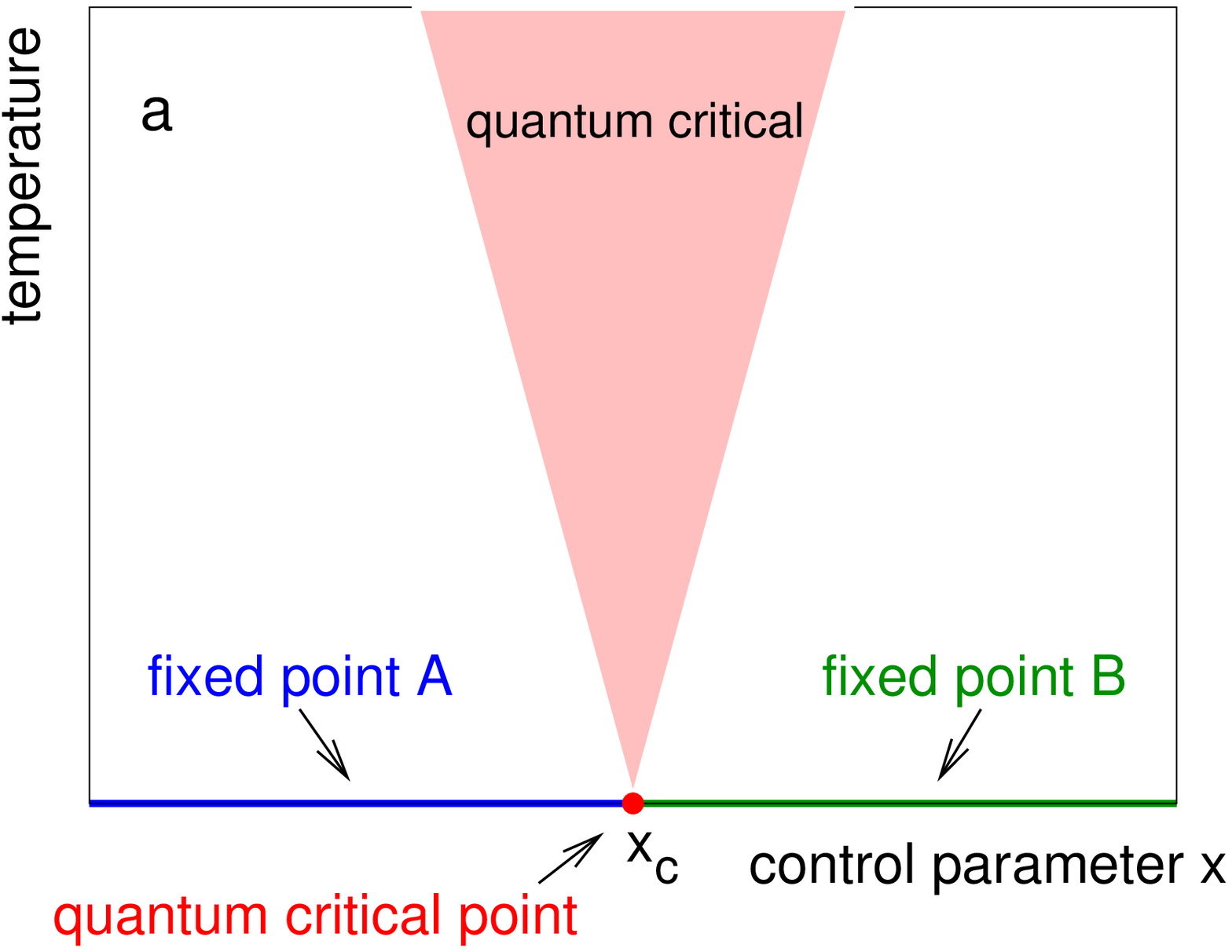}
\includegraphics[width=5.9cm]{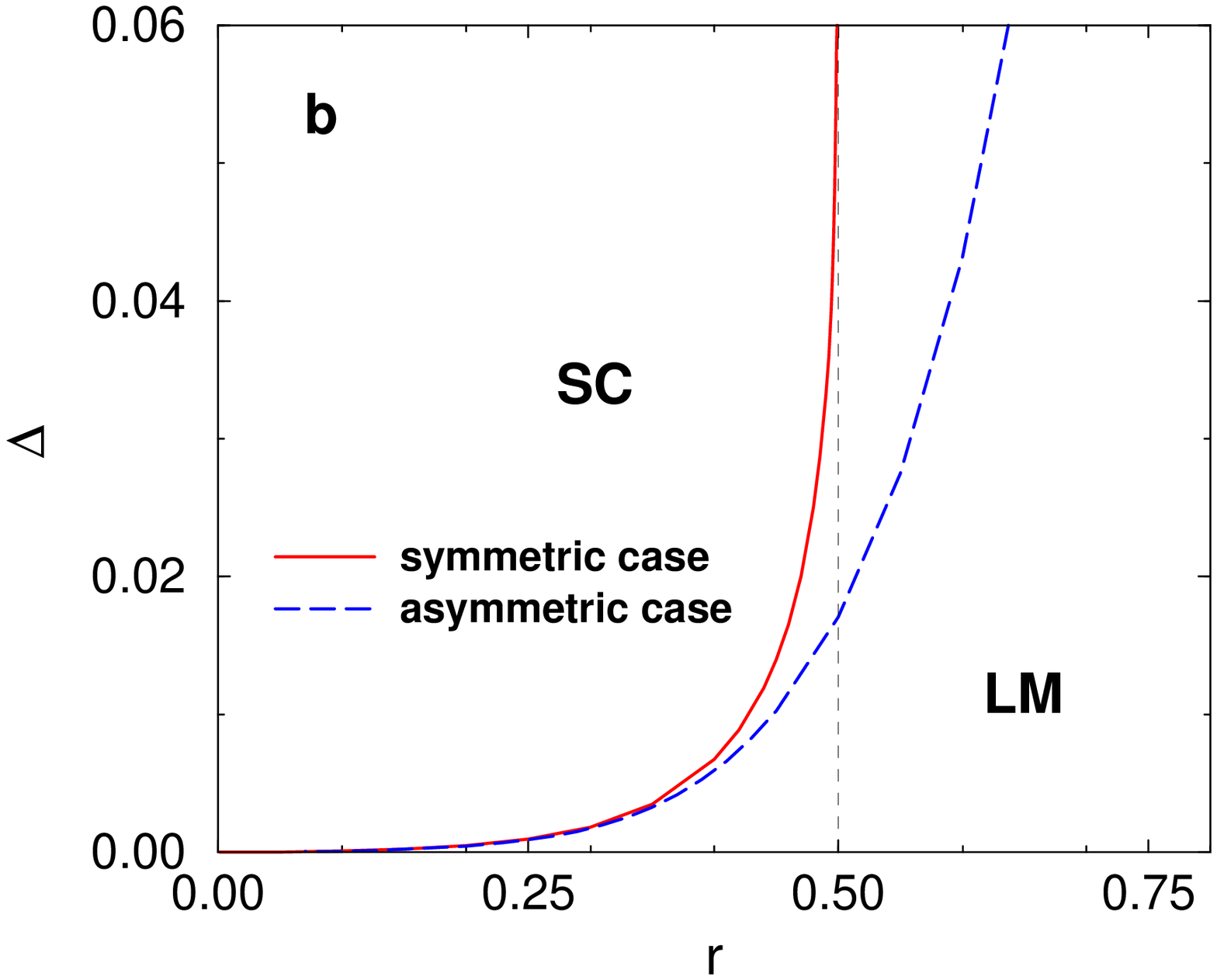}
}
\caption{\small(a) Schematic phase diagram near a quantum phase transition
between fixed points A and B upon variation of a control parameter $x$;
the $T=0$ critical point controls the dynamics in the $T>0$ quantum critical region
whose crossover boundaries are given by $T \sim |x-x_c|^{\nu z}$ where
$\nu$ and $z$ are the correlation length and dynamical critical exponents,
respectively.
         (b) $T=0$ phase diagram for the pseudogap Anderson model
            in the p-h symmetric case (solid line, $U=10^{-3}$,
            $\varepsilon_f = -0.5 \cdot 10^{-3}$, conduction band
            cutoff at -1 and 1) and the p-h asymmetric case
            (dashed line,  $\varepsilon_f = -0.4 \cdot 10^{-3}$);
            $\Delta$ measures the hybridization strength,
            $\widetilde{\Delta}(\epsilon) \equiv \pi V^2 \rho(\epsilon) = \Delta |\epsilon|^r$.
\vspace*{-0.3cm}
}
\end{figure}

The pseudogap case, first considered by Withoff and Fradkin \cite{Wit90},
corresponds to a bath with
$\rho(\varepsilon) \propto |\varepsilon|^r$ ($r>0$), i.e.,
the DOS is zero only {\em at} the Fermi level.
The corresponding Kondo and single-impurity Anderson models interpolate between
the metallic case ($r=0$) and the hard-gap case ($r\to\infty$).
The pseudogap case $0<r<\infty$ leads to a very rich behaviour,
in particular to a continuous transition
between a local-moment and a strong-coupling phase.
Figure 1b shows a typical phase diagram for the pseudogap
Anderson model.
In the p-h symmetric case (solid) the critical coupling $\Delta$,
measuring the hybridization between band electrons and local moment,
diverges at $r=\frac{1}{2}$, and no screening occurs for
$r>\frac{1}{2}$ \cite{GBI,bullapg}.
No divergence occurs for p-h asymmetry (dashed) \cite{GBI}.

We now briefly describe the properties of the fixed points
in the pseudogap Kondo problem \cite{GBI}.
Due to the power-law conduction band DOS, already the stable LM
and SC fixed points show non-trivial behavior \cite{GBI,bullapg}.
The LM phase has the properties of a free spin $\frac{1}{2}$
with residual entropy $S_{\rm imp}=k_B \ln 2$ and
low-temperature impurity susceptibility $\chi_{\rm imp}=1/(4 k_B T)$,
but the leading corrections show $r$-dependent power laws.
The p-h symmetric SC fixed point has very unusual properties,
namely $S_{\rm imp}=2 r k_B \ln 2$, $\chi_{\rm imp}=r/(8 k_B T)$
for $0<r<\frac{1}{2}$.
In contrast, the p-h asymmetric SC fixed point simply displays
a completely screened moment, $S_{\rm imp}= T\chi_{\rm imp}=0$.
The impurity spectral function follows a $\omega^r$ power law
at both the LM and the asymmetric SC fixed point, whereas it
diverges as $\omega^{-r}$ at the symmetric SC fixed point --
this ``peak'' can be viewed as a generalization of the Kondo resonance in
the standard case ($r=0$), and scaling of this peak is observed upon
approaching the SC-LM phase boundary \cite{bullapg,David}.
At the critical point non-trivial behavior corresponding to a fractional moment
can be observed:
$S_{\rm imp}= k_B {\cal C}_S(r)$, $\chi_{\rm imp}= {\cal C}_\chi(r)/(k_B T)$
with ${\cal C}_S$, ${\cal C}_\chi$ being universal functions of $r$ \cite{GBI,MVRB}.
The spectral function displays a $\omega^{-r}$ power law (for $r<1$) with
a remarkable ``pinning'' of the critical exponent.

\begin{figure}
\label{fig2}
\centerline{\includegraphics[width=8.5cm]{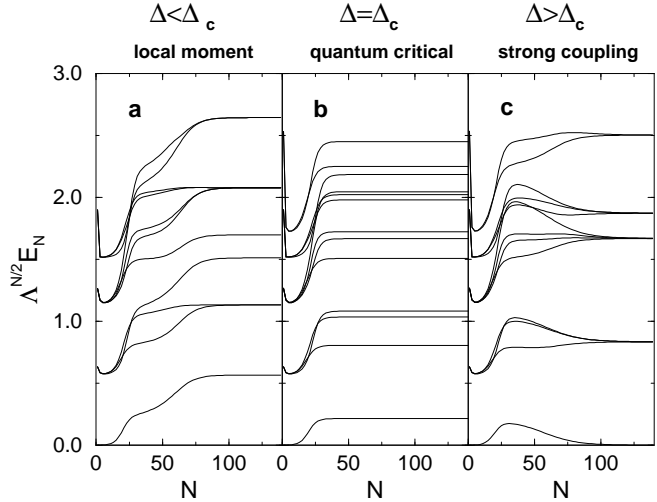}}
\caption{
\small Flow diagrams for the low-energy many-body excitations
obtained from the numerical renormalization group for the three
different fixed points of the soft-gap Anderson model. $N$ is the
number of iterations of the NRG procedure, $\Lambda$ the NRG discretization
parameter.
\vspace*{-0.3cm}
}
\end{figure}

We note that the critical point at small $r$
is perturbatively accessible in a double expansion in
$r$ and $J$.
However, the NRG results suggest that the physics is different
for $r>\frac{1}{2}$; in particular, $r=1$ appears to play
the role of an upper critical ``dimension'' \cite{GBI,MVRB,ingersentsi}.
The universal critical theory of the transitions in the
pseudogap Kondo model is not yet known.
A piece of information, provided by
the numerical renormalization group method, is shown
in Fig. \ref{fig2}. Here we plot the energies of the many-body states
as a function of the iteration number $N$ of the NRG procedure.
Increasing values of $N$ correspond to
decreasing temperature, $N\propto -\ln T$, so we can easily recognize
the three different fixed points for $\Delta < \Delta_{\rm c}$
(Fig. \ref{fig2}a),
$\Delta = \Delta_{\rm c}$ (Fig. \ref{fig2}b), and
$\Delta > \Delta_{\rm c}$ (Fig. \ref{fig2}c).
The structure of the LM and SC fixed points
can be easily understood as that of a free conduction
electron chain \cite{nrg,Kri80}. The combination of the
single-particle states of the free chain leads to the degeneracies
seen in the many-particle states of the LM and SC
fixed points (Fig. \ref{fig2}a and Fig. \ref{fig2}c).
In contrast, the structure of the quantum critical point
is unclear. Degeneracies due to the combination of single-particle
levels are missing, probably because the quantum critical point
is {\em not} build up of non-interacting single-particle states.

The pseudogap Kondo model has been proposed \cite{MVRB} to describe
impurity moments in $d$-wave superconductors ($r=1$), where
signatures of Kondo physics have been found in NMR
experiments \cite{bobroff}.
Furthermore, a large peak seen in STM tunneling near Zn impurities
\cite{seamuszn} can be related to the impurity spectrum
in the asymmetric pseudogap Kondo model.


\section{Multi-channel Kondo model}

Kondo screening is strongly modified if
two or more fermionic screening channels compete.
Nozi\`eres and Blandin \cite{Noz80} proposed a
two-channel generalization of the Kondo model, which shows
overscreening associated with an intermediate-coupling fixed point
and non-Fermi liquid behaviour
in various thermodynamic and transport properties.
In general, such behavior occurs for any number of channels $K>1$
coupled to a spin $\frac{1}{2}$, and does not require fine-tuning of
the Kondo coupling, however, it is unstable w.r.t. a channel 
asymmetry \cite{AJ}.

The anomalous static properties at the two-channel
non-Fermi liquid fixed point are \cite{CZ}
a residual entropy $S_{\rm imp} = \frac{k_B}{2} \ln 2$
(indicating that `half'-fermionic excitations play a crucial role for
the structure of the fixed point),
a logarithmic divergence of the susceptibility $\chi_{\rm imp} \propto \ln T$
and of the specific heat ratio $\gamma = C_{\rm imp}/T \propto \ln T $,
and an anomalous Wilson ratio $R=\chi/\gamma=8/3$, in contrast
to the result for the standard Kondo model $R=8/4=2$.

Analogous to the discussion in the previous section
the many-particle excitation spectrum
of this intermediate-coupling fixed point cannot be decomposed in terms of
usual free fermions \cite{CLN,PC}, however,
a description in terms of non-interacting {\em Majorana} fermions
is possible \cite{BHZ}.
Many of the low-energy properties of the two-channel and related models
have been studied using conformal field theory techniques \cite{AL,olivier}.
Interestingly, the multi-channel Kondo fixed point is perturbatively accessible
in the limit of large channel number ($K\gg 1$) \cite{Noz80,olivier}.
Experimental realizations have been discussed in the context
of rare-earth compounds \cite{CZ}; furthermore, proposals based
on quantum-dot devices have been put forward.


\section{Bose-Fermi Kondo model}

Novel phenomena occur for magnetic impurities coupled
to {\em both} a fermionic and a bosonic bath, where the bosons represent
collective host spin fluctuations.
In the resulting Bose-Fermi Kondo model, the two interactions compete in a
non-trivial manner~\cite{bfk,bfknew}.
Particularly interesting is the case of a bosonic bath with zero or small gap,
corresponding to the vicinity to a magnetic quantum critical point in the
host.
In $(3\!-\!\epsilon)$ dimensions, the bosonic spectral density then follows a
power law $\propto \omega^{1-\epsilon}$.
For $\epsilon>0$, the interaction between the impurity and the bosonic bath can
completely suppress fermionic Kondo screening.
The resulting phase corresponds to an intermediate-coupling fixed point
w.r.t. the impurity--boson interaction.
Here, the impurity moment shows universal fluctuations, with
local spin correlations characterized by a power law,
$\langle S(\tau) S\rangle \propto \tau^{-\epsilon}$,
and a Curie contribution to the impurity susceptibility equivalent
to a fractional spin, $\chi_{\rm imp}= {\widetilde{\cal C}}_\chi(\epsilon)/(k_B T)$~\cite{bfk,bfknew,science}.
On the other hand, large fermionic Kondo coupling leads to a strong-coupling
phase with conventional Kondo screening.

The resulting phase diagram for the Bose-Fermi Kondo model thus shows a Kondo-screened
phase, a bosonic fluctuating phase, and a continuous quantum phase transition in between.
The boundary quantum critical point has magnetic properties similar to
the bosonic fluctuating fixed point \cite{bfknew}.
Both intermediate-coupling fixed points are perturbatively accessible for small $\epsilon$;
it is likely that the structure of the phase diagram also applies to $\epsilon=1$,
however, no accurate numerical calculations are available to date.
Both the critical and the bosonic fluctuating fixed points are
unstable w.r.t. breaking of SU(2) symmetry,
but the structure of the phase diagram is similar for both XY and Ising
symmetries~\cite{bfknew}.

The Bose-Fermi Kondo model has recently received a lot of interest
in the context of the extended dynamical mean-field theory \cite{edmft} where
a lattice model is mapped onto a self-consistent impurity model with both
fermionic and bosonic baths.
In particular, based on neutron scattering experiments on the
heavy-fermion compound $\rm CeCu_{5-x} Au_x$ \cite{schroder},
a self-consistent version of the Bose-Fermi Kondo model has been proposed
to describe a ``local'' critical point in Kondo lattice models.

We further note that the Bose Kondo model with Ising symmetry and an
additional external field is related to the spin-boson model \cite{leggett},
which also shows phase transitions upon variation of the coupling
between spin and dissipative bath.
Spin-boson models have been studied extensively in the context of dissipative
two-level systems, and have applications in many fields like
glass physics, quantum computation etc.


\section{Two-impurity Kondo models}

Models of two impurities offer a new ingredient,
namely the exchange interaction, $I$, between the two
impurity spins which competes with Kondo screening of the
individual impurities.
This inter-impurity interaction, which can lead to a magnetic
ordering transition in lattice models, arises both from direct
exchange and from the Ruderman-Kittel-Kasuya-Yosida (RKKY)
interaction mediated by the conduction electrons.

In the simplest model of two $S=\frac{1}{2}$ impurities,
a ground state singlet ($S_{\rm tot}=0$) can be realized
either by individual Kondo screening (if $I<T_K$)
or by formation of an inter-impurity singlet (if $I>T_K$).
It has been shown that these two parameter regimes are continuously
connected (without a $T=0$ phase transition) as $I$ is varied
in the generic situation without particle-hole symmetry.
Notably, in the particle-hole symmetric case one finds a transition
associated with an unstable non-Fermi liquid fixed
point~\cite{2impnrg,2impsakai,2impcft}.

Quantum phase transitions generically occur in impurity models
showing phases with {\em different} ground state spin.
For two impurities, this can be realized by coupling to a
{\em single} conduction band channel only \cite{VBH}.
In this case, a Kosterlitz-Thouless-type transition between
a singlet and a doublet state occurs, associated with a second
exponentially small energy scale in the Kondo regime \cite{VBH}.
The physics becomes even richer if multi-channel physics
is combined with multi-impurity physics -- here, a variety
of fixed points including such with local non-Fermi liquid
behavior can be realized.

Experimentally, quantum dots provide an ideal laboratory
to study systems of two (or more) ``impurities'' -- note
that the local ``impurity'' states can arise either from
charge or from spin degrees of freedom on each quantum dot.
In particular, a number of experiments have been performed
on coupled quantum dot systems which can be directly mapped onto
models of two Kondo or Anderson impurities \cite{wiel}.
In addition, experimental realizations of two-impurity models
using magnetic adatoms on metallic surfaces appear possible. 


\section{Summary}
We have reviewed a variety of interesting zero-temperature critical points,
which exist in impurity models where conventional Kondo screening
is suppressed by competing physics.
Significant progress has been made in recent years, both in analytical and
numerical work; however, in a number of cases our understanding
concerning, e.g., the nature of the quantum critical points is rather
limited.
Clearly, a further development of theoretical techniques, both perturbative
and non-perturbative, is essential.


It is a pleasure to acknowledge fruitful discussions and collaborations
with A. Hewson, W. Hofstetter, D. Logan, Th. Pruschke, and S. Sachdev,
as well as financial support by the DFG 
through SFB 484.



\begin{thebibliography}{99}


\bibitem{Sachdev} S. Sachdev, {\it Quantum Phase Transitions},
Cambridge University Press, Cambridge (1999).

\bibitem{Hewson}  A.~C. Hewson,
  {\it The Kondo Problem to Heavy Fermions},
  Cambridge University Press, Cambridge (1993).

\bibitem{hardgap}
K. Chen and C. Jayaprakash, Phys. Rev. B {\bf 57}, 5225 (1998).

\bibitem{Wit90}
D. Withoff and E. Fradkin, Phys. Rev. Lett. {\bf 64},  1835 (1990).

\bibitem{GBI}
C.~Gon\-za\-lez-Buxton and K.~Ingersent, Phys. Rev. B {\bf 57}, 14254 (1998).

\bibitem{bullapg}
R.~Bulla, T.~Pruschke, and A.~C.~Hewson, J. Phys.: Condens. Matter {\bf 9}, 10463 (1997);
R.~Bulla, M.~T.~Glossop, D.~E.~Logan, and T.~Pruschke, {\em ibid} {\bf 12}, 4899 (2000).

\bibitem{David} D. E. Logan and M. T. Glossop,
         J. Phys.: Condens. Matter {\bf 12}, 985 (2000).

\bibitem{MVRB} M. Vojta and R. Bulla, Phys. Rev. B {\bf 65}, 014511 (2002).

\bibitem{ingersentsi}
K.~Ingersent and Q.~Si, Phys. Rev. Lett. {\bf 89}, 076403 (2002).


\bibitem{nrg} K.~G.~Wilson, Rev. Mod. Phys. {\bf 47}, 773 (1975).

\bibitem{Kri80}
H.~R. Krishna-murthy, J.~W. Wilkins, and K.~G. Wilson,
Phys. Rev. B {\bf 21}, 1003 (1980).

\bibitem{bobroff}
J.~Bobroff, W.~A.~MacFarlane, H.~Alloul, P.~Mendels,
N.~Blanchard, G.~Collin, and J.-F.~Marucco, Phys. Rev. Lett. {\bf 83}, 4381 (1999).

\bibitem{seamuszn}
E.~W.~Hudson, S.~H.~Pan, A.~K.~Gupta, K.~W.~Ng, and J.~C.~Davis,
Science {\bf 285}, 88 (1999).


\bibitem{Noz80}
P. Nozi\`eres and A. Blandin, J. Physique {\bf 41}, 193 (1980).


\bibitem{AJ} N. Andrei and A. Jerez,
        Phys. Rev. Lett. {\bf 74}, 4507 (1995).

\bibitem{CZ} D. L. Cox and A. Zawadowski, Adv. Phys. {\bf 47}, 599 (1998).

\bibitem{CLN}
  D. M. Cragg, P. Lloyd, and P. Nozi\`eres,
  J. Phys. C, {\bf 13}, 803 (1980).

\bibitem{PC}
   H.-B. Pang and D. L. Cox, Phys. Rev. B {\bf 44}, 9454 (1991).

\bibitem{BHZ}
R. Bulla, A. C. Hewson, and G.-M. Zhang, Phys. Rev. B {\bf 56}, 11721 (1997).

\bibitem{AL}
I. Affleck and A. W. W. Ludwig,
Nucl. Phys. B {\bf 352}, 849 (1991) and {\bf 360}, 641 (1991),
Phys. Rev. B {\bf 48}, 7297 (1993).

\bibitem{olivier}
O. Parcollet, A. Georges, G. Kotliar, and A. Sengupta,
Phys. Rev. B {\bf 58}, 3794 (1998).


\bibitem{bfk}
J.~L.~Smith and Q.~Si, cond-mat/9705140, Europhys. Lett. {\bf 45}, 228 (1999);
A.~M.~Sengupta, Phys. Rev. B {\bf 61}, 4041 (2000).

\bibitem{bfknew}
L. Zhu and Q. Si, Phys. Rev. B {\bf 66}, 024426 (2002);
G. Zarand and E. Demler, Phys. Rev. B {\bf 66}, 024427 (2002).

\bibitem{science}
S. Sachdev, C. Buragohain, and M. Vojta, Science {\bf 286}, 2479 (1999);
M. Vojta, C. Buragohain, and S. Sachdev, Phys. Rev. B {\bf 61}, 15152 (2000).

\bibitem{edmft}
Q.~Si, S.~Rabello, K.~Ingersent, and J.~L.~Smith, Nature {\bf 413}, 804 (2001).

\bibitem{schroder}
A. Schr\"oder, G. Aeppli, R. Coldea, M.  Adams,
O. Stockert, H.~v.~L\"ohneysen, E. Bucher, R.~Ramazashvili, and P. Coleman,
Nature {\bf  407}, 351 (2000).

\bibitem{leggett}
A. J. Leggett, S. Chakravarty, A. T. Dorsey, M. P. A. Fisher, A. Garg,
and W. Zwerger, Rev. Mod. Phys. {\bf 59}, 1 (1987).



\bibitem{2impnrg}
B.~A.~Jones and C.~M.~Varma, Phys. Rev. Lett. {\bf 58}, 843 (1987);
B.~A.~Jones, C.~M.~Varma, and J.~W.~Wilkins, {\em ibid.} {\bf 61}, 125 (1988).

\bibitem{2impsakai}
O.~Sakai, Y.~Shimizu, and T.~Kasuya, Solid State Comm. {\bf 75}, 81 (1990);
O.~Sakai and Y.~Shimizu, J. Phys. Soc. Jpn {\bf 61}, 2333 (1992),
{\em ibid}, {\bf 61}, 2348 (1992).

\bibitem{2impcft}
I.~Affleck, A.~W.~W.~Ludwig, and B.~A.~Jones, Phys. Rev. B {\bf 52}, 9528 (1995).


\bibitem{VBH}
M. Vojta, R. Bulla, and W. Hofstetter, Phys. Rev. B {\bf 65}, 140405(R) (2002).










\bibitem{wiel}
W.~G. van der Wiel, S. De Franceschi, J. M. Elzerman, S. Tarucha, L. P. Kouwenhoven,
J. Motohisa, F. Nakajima, and T. Fukui,
Phys. Rev. Lett. {\bf 88}, 126803 (2002).



\end{thebibliography}
\end{document}